# Understanding the Origins of a Pandemic Virus

Carlos Xavier Hernández[1], Joseph Chan[1], Hossein Khiabanian[1,2], Raul Rabadan[1,2*]

[1] Center for Computational Biology and Bioinformatics,

[2] Department of Biomedical Informatics,

Columbia University College of Physicians and Surgeons,

New York, NY, United States

* Corresponding author's email: rabadan@dbmi.columbia.edu

**Abstract**

Understanding the origin of infectious diseases provides scientifically based rationales for implementing public health measures that may help to avoid or mitigate future epidemics. The recent ancestors of a pandemic virus provide invaluable information about the set of minimal genomic alterations that transformed a zoonotic agent into a full human pandemic. Since the first confirmed cases of the H1N1 pandemic virus in the spring of 2009 [1,2], several hypotheses about the strain's origins have been proposed. However, how, where, and when it first infected humans is still far from clear. The only way to piece together this epidemiological puzzle relies on the collective effort of the international scientific community to increase genomic sequencing of influenza isolates, especially ones collected in the months prior to the origin of the pandemic.

The H1N1 pandemic in 2009 was the first influenza pandemic of the 21st century, a period in which technological developments have allowed scientists around the globe to isolate and sequence viral genomes and share results in almost real time. These advances present an opportunity to learn how influenza pandemics start, spread and evolve. Despite all the worldwide effort, there are many questions that remain unanswered about the origin of the 2009 H1N1 influenza virus (H1N1pdm). The closest ancestors of the pandemic virus were related to viruses isolated in swine more than a decade ago [3,4], but it is still unclear how, where, and when it became a human pandemic, and where the virus was circulating the year before the pandemic was declared. The most accepted conjecture is the "hidden pig herd hypothesis," which suggests that incomplete surveillance missed strains circulating in unsampled swine herds [3,4].

Since March 2009, when the first pandemic cases were reported, until May 1st, 2009, hundreds of H1N1pdm isolates were collected and sequenced, all of which came from human hosts except for one isolate from a pig in Mexico (A/swine/4/Mexico/2009 (H1N1))[1]. After May 1st, 2009, more than one hundred H1N1pdm virus sequences isolated in swine have been deposited in the National Center for Biotechnology Information Influenza Virus Resource[7], all of which could be traced back to human H1N1pdm viruses, hinting at a possible human to swine transmission, with no evidence of swine to human transmission. Although, GenBank also contains more than a hundred swine influenza genomes, isolated from three continents prior the start of the pandemic (between 2008 and the spring of 2009), none seem directly related to the H1N1pdm pandemic virus. It is indeed surprising that only one H1N1pdm swine isolate is publically known before the pandemic virus was widely spread in the human population.

---

[1] As of April 18th, 2011, 413 H1N1pdm hemagglutinin sequences were deposited in GenBank[7] from isolates collected before May 1st, 2009, of a total of 4,455 H1N1pdm isolates deposited. Similar isolate numbers could be found in GIAID[8].

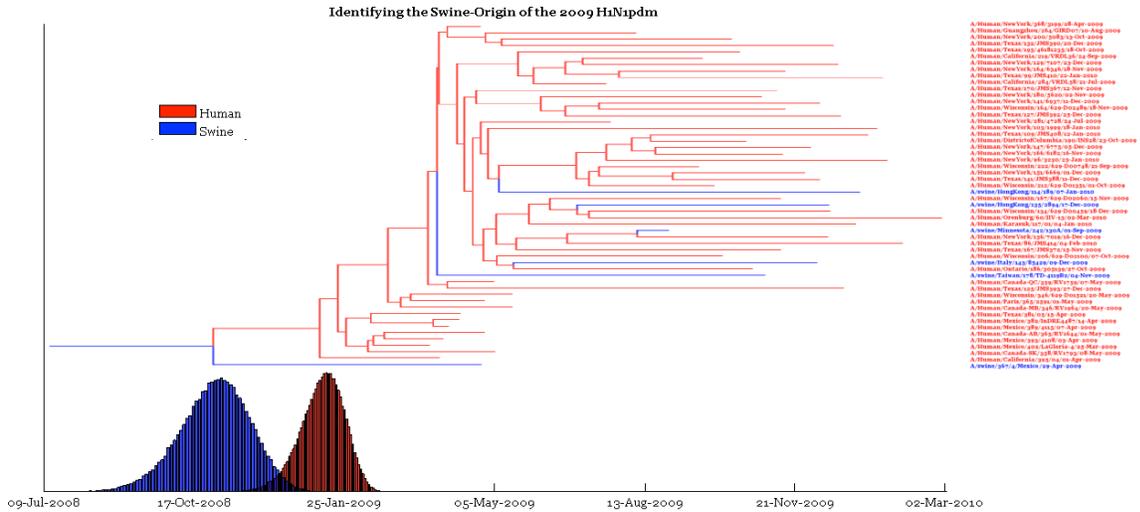

**Figure 1:** Bayesian Markov chain Monte Carlo analysis of 54 human (red) and swine (blue) influenza sequences. The most recent common ancestor of all the strains dates back to the last weeks of 2008, prior the date of the most recent common ancestor of the human pandemic strains.

It may be impossible to verify the exact time of the first H1N1pdm infection in humans; however, the time to the most recent common ancestor (TMRCA) of all human H1N1pdm strains can be estimated to be in the first two months of 2009 [5]. Bayesian Markov chain Monte Carlo (MCMC) methods provide an excellent tool for such estimates [6]. Figure 1 represents the distribution of the TMRCA of 54 early H1N1pdm isolates using Bayesian MCMC analysis (see Methods). In red is the distribution of the dates of the TMRCA for the human strains, identical to the results of Rambaut and Holmes [5]. Interestingly, when the earliest swine isolate (A/swine/4/Mexico/2009 (H1N1)) that is related to the H1N1pdm is included, the TMRCA retraces back to mid-November of 2008, suggesting that there could be a common ancestor to the swine isolate and the most recent common ancestor of the human isolates (Swine TMRCA: October 31st, 2008 +/- 23 days, Human TMRCA: January 11th, 2009 +/- 15 days). Unfortunately, this interesting isolate is poorly annotated (GenBank information: swine in Queretaro, Mexico, in April 2009), without accompanying publication.

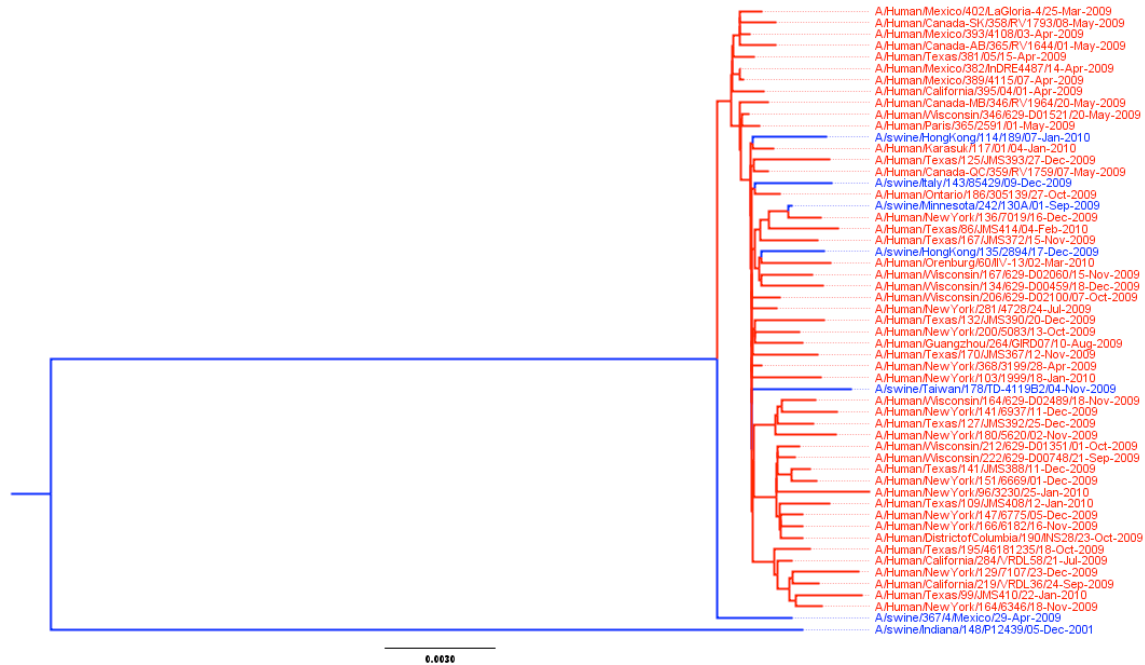

**Figure 2:** When rooted with a common ancestor (A/swine/Indiana/P12439/00 (H1N2)) for concatenated segments PB2, PB1, PA, HA, NP and NS, Maximum Likelihood and Neighbor Joining trees show similar results (NJ tree shown here): A/swine/4/Mexico/2009 (H1N1) branches off before human H1N1pdm isolates.

Alternatively, similar results are obtained by rooting the phylogenetic tree with an ancestor of the pandemic virus. We chose the strain A/Swine/Indiana/P12439/00 (H1N2), a swine isolate from the year 2000 that has 95% similarity with the pandemic virus in all segments apart from NA and MP. In the Maximum Likelihood (ML) and Neighbor Joining (NJ) trees (Figure 2), all human isolates branch out together, indicating a common recent origin to the human pandemic, similar to the previous analysis (bootstrap values of A/swine/4/Mexico/2009 (H1N1) isolate branch between 80 and 97). However, the oldest swine isolate branches out independently, but remains nearby, indicating a related, yet independent, older common ancestor to the human pandemic. To rule out the possibility that an H1N1pdm isolate, not present in the selected sample of sequences, could have altered our conclusions, we confirmed the robustness of our results by extending analysis to 166 earliest H1N1pdm genomes isolated from March to May 1st 2009, and to random subsamples of 1,843 H1N1pdm genomes (see Methods). Changing the outgroup to another swine isolate reported[4] to be related to the ancestors of the pandemic (A/swine/Hong Kong/915/2004 (H1N2)) did not change the results.

Table 1

| Segment | CDSPosition | Human H1N1pdm | | Mexico/swine H1N1 | | Indiana H1N2 2000 Hong Kong H1N2 2004 | |
|---|---|---|---|---|---|---|---|
| | | nt | aa | nt | aa | nt | aa |
| PB1 | 1722 | T | F | C | F | C | F |
| | 1803 | A | I | C | I | C | I |
| NP | 159 | T | D | G | E | G | E |
| | 792 | C | L | T | L | T | L |
| NS1 | 408 | A | V | G | V | G | V |

Five sites are common to A/Swine/Indiana/P12439/00 (H1N2) and A/swine/4/Mexico/2009 (H1N1) swine isolates and different from the other early human H1N1pdm isolates (any given early human isolate showed less homology to the ancestor from Indiana in sites not common to the other early isolates). All sites are 3$^{rd}$ codon positions, 4 synonymous and 1 non-synonymous in NP (conservative change Glutamic to Aspartic Acid). These positions were confirmed using A/swine/Hong Kong/915/2004 (H1N2) as an outgroup (Table 1).

Several caveats need to be considered in this analysis, and the results must be interpreted with caution. Poor sequencing of one isolate could artificially elongate a branch, displacing the TMRCA to an earlier date. We performed the analysis with different clustering techniques and introduced artificial errors into the sequences. Although the branch lengths in the ML and NJ trees vary, their topology does not change. Therefore, possible poor sequencing does not sufficiently explain why the swine isolate branches off before all human isolates in the rooted trees. The trees were constructed by concatenating all segments except NA and MP. If the swine isolate was a reassortant with a different segment structure to other H1N1pdm genomes, the information from the different segments could interfere to produce earlier branches. However, the pattern of expected nucleotide substitutions is consistent within each segment, suggesting that the swine isolate has the same segment structure as the other H1N1pdm genomes.

Although the previous analyses suggest that the pre-pandemic virus could have been circulating in Mexican swine, we want to emphasize that strong conclusions about the origin of the pandemic could not and should not be drawn from a single isolate. Possible concerns include laboratory contaminations and mislabeling, both of which are hard to rule out. Therefore, only careful re-analysis of the swine isolate and/or analysis of other similar isolates could create a clearer picture of the pre-pandemic viral landscape.

The purpose of this note is to encourage the scientific community to continue the efforts to investigate the origins of this pandemic. Many other factors, especially those politically and economically sensitive, might play a role in the public release of pre-pandemic sequence data. However, understanding the origins of a pandemic should not be written off as a mere academic exercise. Influenza pandemics have arisen every 30 years, and without a clear understanding of pre-pandemic conditions, they will certainly recur. Hence, encouraging disclosure of scientific information, beyond non-scientific concerns is essential for implementing effective measures to avoid future outbreaks.

**Methods**

H1N1pdm sequences were collected from Influenza Virus Resource of the National Center for Biotechnology Information (NCBI)[7]. For the Bayesian Markov chain Monte Carlo analysis, human H1N1pdm sequences were filtered to produce a dataset that met the following criteria: that the complete genome of the virus was available, without unidentified bases and that the exact date of collection was recorded. A further filtering was applied in which, for each day of isolation, only one virus from a given location (country or state, if applicable) was retained. This resulted in a total of 54 isolates (Appendix 1). All segments were concatenated. Swine isolates were also collected from the NCBI and filtered for complete genomes.

The NJ analysis was extended to 166 early H1N1pdm viruses isolated before May 1st, 2009 (segments PB2, PB1, PA, HA, NP and NS, see Appendix 2). Bootstrap value of A/swine/4/Mexico/2009 (H1N1) branch separated from all human isolates was 82%. To ensure that the observed topology in the rooted phylogenetic tree did not arise by chance or a relevant isolate was left out of the analysis, we iteratively sampled 52 sequences from 1,843 human H1N1pdm sequences (concatenated PB2, PB1, PA, HA and NP segments). The complete list of 1,843 names can be requested from the authors. In each iteration, we combined these random human sequences with A/swine/4/Mexico/2009 (H1N1) and

A/Swine/Indiana/P12439/00 (H1N2), and created 100 phylogenetic trees by maximum likelihood using Phylip[10] with default parameters. 97 of the resulting trees showed that all randomly sampled human sequences collectively branched from the rooted Indiana swine ancestor and the Mexican swine isolate. Bootstrap values with 20 replicates were calculated at this junction for all trees with a mean of 70.1, a standard deviation of 11.43, and a range of [45,95]. 3 of the 100 trees in which the Mexican swine isolate failed to split first with the entire human clade contained low bootstrap values (<45%).

Eight bases in A/swine/4/Mexico/2009 (H1N1) appear partially unresolved (3 R in HA, 2 K in PB2 and M1, 1 M in NA, 2 Y in NA and M1) in conserved sites relative to other H1N1pdm genomes. Partially unresolved bases can be corrected to the respective values in one of the earlier H1N1pdm sequence (A/California/04/2009 (H1N1)), without affecting the results of the analyses in this manuscript. Poor quality in sequence process could introduce substitutions; substitutions that would appear in a random fashion in both synonymous and non-synonymous sites. Most of the substitutions observed between A/swine/4/Mexico/2009 (H1N1) and a representative of human H1N1pdm virus (A/California/04/2009 (H1N1)) are synonymous (28 of 41), including the four common substitutions with A/swine/Hong Kong/915/2004 and A/Swine/Indiana/P12439/00 (H1N2) described in the main text. These observations suggest a minor role of sequence errors in the analyses. Dnds ratios[9] between A/swine/4/Mexico/2009 (H1N1) and other H1N1pdm isolates is 0.144, similar to other H1N1pdm pairs: 0.1367 [95% CI: 0.0646, 0.2900] (distribution from the 1,843 human H1N1pdm isolates).

Sequences were analyzed using BEAST v1.6.0 [5,6], a Bayesian Markov chain Monte Carlo (MCMC) approach to sample time-structured evolutionary trees from their joint posterior probability distribution. The dataset was analyzed under an exponential-growth coalescent model as a prior on the tree, the HKY+ $\Gamma$ model of nucleotide substitution. Relaxed and strict molecular clocks were employed, with both yielding similar results (relaxed and strict both provide the same result in 54 isolates, 0.0032 nucleotide/year, and date of human TMRCA Jan. 12$^{th}$, 2009, similar to previous estimates [4,5]). Three independent runs of one hundred million steps each were performed, compared for convergence and combined with less than

10% burn-in from each[5]. The consistency of the procedure was evaluated in the 166 early isolates (Appendix 2)(strict clock, evolutionary rate 0.0038 nucl/year, date of human TMRCA Jan. 18th, 2009), however, the evolutionary rate estimates presented higher variance, probably due to the limited sampling time frame.

**Acknowledgements.** We would like to thank Ben Greenbaum, Eddie Holmes, Arnold Levine, Ian Lipkin, David Lipman, Steve Morse, Martha Nelson, Gustavo Palacios, Vincent Racaniello, and Vladimir Trifonov, for helpful discussions and comments. The work of the authors is supported by the Northeast Biodefence Center (U54-AI057158), the National Institute of Health (U54 CA121852-05), and the National Library of Medicine (1R01LM010140-01).

**APPENDIX 1: Selected isolates.**

A/Human/California/04/01-Apr-2009
A/Human/California/VRDL36/24-Sep-2009
A/Human/California/VRDL58/21-Jul-2009
A/Human/Canada-AB/RV1644/01-May-2009
A/Human/Canada-MB/RV1964/20-May-2009
A/Human/Canada-QC/RV1759/07-May-2009
A/Human/Canada-SK/RV1793/08-May-2009
A/Human/DistrictofColumbia/INS28/23-Oct-2009
A/Human/Guangzhou/GIRD07/10-Aug-2009
A/Human/Karasuk/01/04-Jan-2010
A/Human/Mexico/4108/03-Apr-2009
A/Human/Mexico/4115/07-Apr-2009
A/Human/Mexico/InDRE4487/14-Apr-2009
A/Human/Mexico/LaGloria-4/25-Mar-2009
A/Human/NewYork/1999/18-Jan-2010
A/Human/NewYork/3199/28-Apr-2009
A/Human/NewYork/3230/25-Jan-2010
A/Human/NewYork/4728/24-Jul-2009
A/Human/NewYork/5083/13-Oct-2009
A/Human/NewYork/5620/02-Nov-2009
A/Human/NewYork/6182/16-Nov-2009
A/Human/NewYork/6346/18-Nov-2009
A/Human/NewYork/6669/01-Dec-2009
A/Human/NewYork/6775/05-Dec-2009
A/Human/NewYork/6937/11-Dec-2009
A/Human/NewYork/7019/16-Dec-2009
A/Human/NewYork/7107/23-Dec-2009
A/Human/Ontario/305139/27-Oct-2009
A/Human/Orenburg/IIV-13/02-Mar-2010
A/Human/Paris/2591/01-May-2009
A/Human/Texas/05/15-Apr-2009
A/Human/Texas/46181235/18-Oct-2009
A/Human/Texas/JMS367/12-Nov-2009
A/Human/Texas/JMS372/15-Nov-2009
A/Human/Texas/JMS388/11-Dec-2009
A/Human/Texas/JMS390/20-Dec-2009
A/Human/Texas/JMS392/25-Dec-2009
A/Human/Texas/JMS393/27-Dec-2009
A/Human/Texas/JMS408/12-Jan-2010
A/Human/Texas/JMS410/22-Jan-2010
A/Human/Texas/JMS414/04-Feb-2010
A/Human/Wisconsin/629-D00459/18-Dec-2009
A/Human/Wisconsin/629-D00748/21-Sep-2009
A/Human/Wisconsin/629-D01351/01-Oct-2009

A/Human/Wisconsin/629-D01521/20-May-2009
A/Human/Wisconsin/629-D02060/15-Nov-2009
A/Human/Wisconsin/629-D02100/07-Oct-2009
A/Human/Wisconsin/629-D02489/18-Nov-2009
A/swine/4/Mexico/29-Apr-2009
A/swine/Indiana/P12439/05-Dec-2001
A/swine/Italy/85429/09-Dec-2009
A/swine/HongKong/189/07-Jan-2010
A/swine/HongKong/2894/17-Dec-2009
A/swine/Taiwan/TD-4119B2/04-Nov-2009
A/swine/Minnesota/130A/01-Sep-2009

## APPENDIX 2: Early isolates (<May 1st 2009) with complete genome.

A/Iowa/04/2009
A/Paris/2580/2009
A/MexicoCity/020/2009
A/SanAntonio/PR922/2009
A/Israel/276/2009
A/NewYork/3100/2009
A/NewYork/3217/2009
A/NewYork/3194/2009
A/Brawley/40081/2009
A/Texas/08/2009
A/NewYork/3012/2009
A/NewYork/3348/2009
A/Mexico/InDRE4487/2009
A/NewYork/22/2009
A/NewYork/15/2009
A/Florida/04/2009
A/NewYork/13/2009
A/NewYork/3002/2009
A/NewYork/3206/2009
A/SanAntonio/PR921/2009
A/Canada-ON/RV1590/2009
A/NewYork/21/2009
A/NewYork/3236/2009
A/California/04/2009
A/Canada-NS/RV1582/2009
A/Indiana/09/2009
A/SanDiego/WR1629P/2009
A/Canada-QC/RV1595/2009
A/Mexico/4604/2009
A/Wisconsin/629-D00447/2009
A/Wisconsin/629-D01505/2009
A/MexicoCity/019/2009

A/NewYork/3074/2009
A/NewYork/3262/2009
A/RhodeIsland/02/2009
A/Arizona/02/2009
A/NewYork/3227/2009
A/Alabama/02/2009
A/NewYork/3230/2009
A/NewYork/3166/2009
A/Canada-NS/RV1565/2009
A/NewYork/3168/2009
A/NewYork/3181/2009
A/NewYork/1682/2009
A/NewYork/3198/2009
A/NewYork/18/2009
A/Canada-PQ/RV1602/2009
A/HongKong/415742/2009
A/NewYork/3237/2009
A/NewYork/1669/2009
A/Brawley/WR1636P/2009
A/NewYork/3172/2009
A/Texas/15/2009
A/MexicoCity/024/2009
A/Wisconsin/629-D01558/2009
A/Canada-NS/RV1559/2009
A/NewYork/3234/2009
A/NewYork/3186/2009
A/NewYork/3210/2009
A/Canada-NS/RV1561/2009
A/Maryland/05/2009
A/Netherlands/602/2009
A/SanDiego/WRAIR1648P/2009
A/NewYork/3124/2009
A/Canada-ON/RV1545/2009
A/NewYork/3232/2009
A/NewYork/09/2009
A/Mexico/LaGloria-3/2009
A/NewYork/3176/2009
A/NewYork/3215/2009
A/NewYork/3225/2009
A/Mexico/48N/2009
A/NewYork/3014/2009
A/NewYork/31/2009
A/Canada-NB/RV1546/2009
A/Canada-BC/RV1719/2009
A/CherryPoint/WR1355/2009
A/NewYork/3196/2009

A/Kansas/03/2009
A/Canada-AB/RV1531/2009
A/swine/4/Mexico/2009
A/SouthCarolina/09/2009
A/NewYork/3211/2009
A/SanDiego/WR1631P/2009
A/Canada-NS/RV1554/2009
A/Canada-PQ/RV1586/2009
A/NewYork/3203/2009
A/NewYork/3184/2009
A/Canada-SK/RV1683/2009
A/Mexico/47N/2009
A/NewYork/3195/2009
A/Tennessee/05/2009
A/Colorado/03/2009
A/Mexico/LaGloria-4/2009
A/Massachusetts/06/2009
A/Minnesota/02/2009
A/Mexico/InDRE13555/2009
A/Ohio/07/2009
A/Mexico/InDRE13494/2009
A/California/14/2009
A/NewYork/3365/2009
A/Paris/2590/2009
A/Canada-NS/RV1562/2009
A/Texas/05/2009
A/NewYork/3008/2009
A/Michigan/02/2009
A/NewYork/3191/2009
A/NewYork/3173/2009
A/Texas/22/2009
A/Brawley/40082/2009
A/Canada-BC/RV1720/2009
A/Wisconsin/629-D02361/2009
A/Mexico/InDRE13551/2009
A/Canada-NS/RV1535/2009
A/Mexico/4269/2009
A/Wisconsin/629-D01773/2009
A/Canada-NS/RV1552/2009
A/NewYork/20/2009
A/Wisconsin/629-D00750/2009
A/Arizona/01/2009
A/NewYork/3214/2009
A/Canada-ON/RV1527/2009
A/NewYork/3265/2009
A/SanAntonio/PR923/2009

A/NewYork/3007/2009
A/NewYork/3209/2009
A/Canada-NS/RV1551/2009
A/CherryPoint/WR0080/2009
A/Wisconsin/629-D01529/2009
A/Wisconsin/629-D00767/2009
A/Canada-BC/RV1718/2009
A/Wisconsin/629-D00015/2009
A/NewYork/3189/2009
A/Mexico/InDRE13547/2009
A/NewYork/19/2009
A/Houston/13OS/2009
A/NewYork/10/2009
A/NewYork/3199/2009
A/NewYork/3167/2009
A/BritishColumbia/GFA0401/2009
A/Canada-SK/RV1788/2009
A/Paris/2573/2009
A/Wisconsin/629-D02141/2009
A/NewYork/3247/2009
A/Mexico/InDRE13495/2009
A/Kansas/02/2009
A/NewYork/3193/2009
A/NewYork/3204/2009
A/NewYork/3235/2009
A/Texas/09/2009
A/NewYork/3188/2009
A/California/06/2009
A/Canada-NS/RV1560/2009
A/NewYork/11/2009
A/NewYork/3260/2009
A/Canada/GFA0402/2009
A/NewYork/23/2009
A/Mexico/LaGloria-8/2009
A/Canada-NS/RV1572/2009
A/Mexico/4115/2009
A/Mexico/4486/2009
A/Brawley/WR1635P/2009
A/NewYork/06/2009
A/Toronto/3141/2009
A/England/195/2009
A/Canada-ON/RV1589/2009
A/swine/Indiana/148/P12439/05-Dec-2001